# Electric Control of Multiple Domain Walls in Pt/Co/Pt Nanotrack with Perpendicular Magnetic Anisotropy

Kab-Jin Kim[1], Jae-Chul Lee[1,2], Sang-Jun Yun[1], Gi-Hong Gim[1], Kang-Soo Lee[1], Sug-Bong Choe[1,a)] and Kyung-Ho Shin[2,b)]

[1]Department of Physics, Seoul National University, Seoul 151-742, Republic of Korea

[2]Center for Spintronics Research, Korea Institute of Science and Technology, Seoul 136-791, Republic of Korea

a) Electronic mail: sugbong@snu.ac.kr

b) Electronic mail: kshin@kist.re.kr

Electric control of multiple domain walls (DWs) motion is demonstrated by Pt/Co/Pt nanotracks with perpendicular magnetic anisotropy. Due to the weak microstructural disorders with small DW propagation field, the purely current-driven DW motion is achieved in the creep regime at current densities less than $10^7$ A/cm$^2$ at room temperature. It is confirmed that by use of a scanning magneto-optical Kerr effect microscope, several DWs are simultaneously and identically displaced by the same distance in the same direction. Utilizing such DWs motion, we succeed to realize random bits writing and transferring as a device prototype of four-bit shift registers.

In modern digital magnetic devices, information bits '1' or '0' are stored in the form of magnetic domains with magnetization pointing 'up' or 'down'. Such magnetic bits have been controlled by the conventional magnetic-field-based techniques. However, the recent demonstration on spin-current-based technique[1,2] has opened up great opportunities for novel device concepts with multiple DW shift in a direction contingent upon the current pulses.[3] The operation principles have been verified in several materials such as metallic Permalloy[4] and semiconducting (Ga,Mn)As.[5] However, there remain several urgent issues yet, including how to lower the operation power for Permalloy and/or to increase the Curie temperature for (Ga,Mn)As. Recently, the use of metallic ferromagnets with perpendicular magnetic anisotropy (PMA) has been proposed,[6-12] motivated by the predictions of low operation current density $J_{OP}$,[6] high efficiency,[7,8] and small DW size.[9] They also exhibit simple DW structures (Bloch or Néel types), in contrast to the complex DW structures (vortex and transverse types) in Permalloy nanowires. Despite these appealing predictions, it lacks experimental verification. This is primarily due to the microstructural disorders in these materials, which result in considerable DW pinning,[10] and consequently, high $J_{OP}$ for DW motion.[11,12]

The present work aims to achieve low $J_{OP}$ by reducing the microstructural disorders in metallic PMA materials. For the study, we prepared 5.0-nm Ta/2.5-nm Pt/0.3-nm

Co/1.5-nm Pt films on Si/100-nm SiO$_2$ substrates by dc-magnetron sputtering. The thicknesses of each layer are optimized for minimum microstructural disorders without ruining PMA. The films exhibit a strong PMA of $(8.3\pm0.2)\times10^5$ J/m$^3$ with the saturation magnetization of 1.6±0.2 T, measured by angle-dependent extraordinary Hall effect measurement[13] and alternating gradient magnetometry. The DW in the films propagates circularly with little jaggedness as shown in Fig. 1(a). Such DW propagation is attained even at a small magnetic field, about a few mT. These observations manifest weak microstructural disorders in the films. 500 nm-wide and 16 μm-long nanotracks as shown in Fig. 1(b) are then patterned into the films by use of electron beam lithography and ion milling. To inject current pulses into the nanotracks, two electrodes with coplanar waveguide geometry are stacked on both ends of nanotracks.

Figure 1(c) shows the schematic diagram of experimental procedure. We first saturate the magnetization to the 'down' direction (blue arrow) by applying a magnetic field pulse (150 Oe, 0.5 s). Then, we inject a current pulse $I_R$ (20 mA, 1 μs) through the current line from the function generator 1 (FG1) to the ground. Consequently, a reversed 'up' domain (red arrow) is recorded by the Oersted field (purple arrow) on the right side of the current line. Here, we define the logical bits '1' and '0' as the magnetic domains points 'up' and 'down', respectively. Thus, the bit '1' is recorded by the positive current through the current line, whereas the bit '0' is recorded by the negative current. By injecting a current pulse $I_T$ from the function generator 2 (FG2) to the ground, the DWs—located between two different magnetic domains—are pushed to a side via the spin transfer torque (STT).[1,2] Based on these procedures, we can record and transfer the random magnetic bits in the nanotracks. The magnetic bits are then detected by means of a scanning magneto-optical Kerr effect (MOKE) microscope.[14] The MOKE signal has two distinct levels corresponding to 'up' and 'down' domains, where the higher signal level corresponds to the bit '1'. Two different detection schemes are adopted: 1) scanning along the nanotracks for DW position detection and 2) monitoring at a fixed position for detection of DW passage in time. Note that the present optical detection technique, in contrast to the resistance measurement schemes,[3,4,11] can readily resolve the positions of multiple DWs.

Figure 2(a) shows the purely current-driven DW shift at zero magnetic field bias. The line-scanned MOKE signals along the nanotrack are shown for several successive current pulses $I_T$ ($\pm 9.8 \times 10^6$ A/cm$^2$, 20 ms).[15] The DW shift is caused by the DW creep as magnetic-field-driven DW motion does,[14] which is governed by the thermally activated process overcoming the depinning energy barrier at a magnetic field (or current) smaller than the critical value. The polarity of $I_T$ is denoted in the figure. A single DW (left) and double DWs (right) are tested. In the plots, the MOKE signal is normalized between the two background signals, preliminarily obtained for the 'up' and 'down' saturated states.[16] As the right panel clearly illustrates, the two DWs are transferred to the same direction, defined by the polarity of the electric current pulses. In addition, the speed of the DWs is the same, irrespective to the DW types (i.e. the magnetization of the adjacent domains), which excludes the possibility of the Oersted-field-driven effect.

The parallel shift of the multiple DWs manifests that the shift is ascribed to the STT, which causes the DW shift in the direction determined by $P \times J_{OP}$. Here, the spin current polarization $P$ in our ultrathin Co film interfaced by Pt layers is predicted to be negative,[17] which thus induces the DW motion in the direction of current, in contrast to the direction of electron flow in previous reports.[4,5] Several other possible mechanisms such as the the hydromagnetic DW drag,[18,19] the Hall charge drag,[19] and the Rashba effect[20] are also examined, but found to be irrelevant to our observation: The hydromagnetic and Hall charge effects in our sample geometry are estimated to be several orders smaller than our experimental observation. The Rashba effect is negligible in our (almost) symmetric layer structure.

A shift register[4,21] is then built up with the magnetic nanotrack. Utilizing the current-driven DW motion at such a low current density, serial-in unidirectional shift register is readily realized. To demonstrate a 4-bits shift register, we first determine the optimal duration (~50 ms) of the current pulses $I_T$ ($9.8\times10^6$ A/cm$^2$) to transfer DWs for a quarter of the nanotrack length. The current pulse is then used as the clock pulse for the shift register. To diagnose the device operation, random bit sequences are recorded for each clock and checked to see whether the data bits are transferred accordingly. Figure 2(b) shows the line-scanned MOKE signal for each clock. The figure clearly shows that the recorded data sequences 1010 (left) and 1100 (right) are transferred exactly a quarter of the device length for each clock pulse, resulting in 4 individual sections of serial magnetic bits. These results manifest that the multiple DWs with random sequence can be reliably controlled by the electric current pulses.

Figure 3 depicts the results of the operation timetable for the 4-bit shift register. The top panel specifies the current pulses $I_R$ to record the arbitrary bit sequence (here, 00111011 repeatedly). The second panel indicates the clock pulses $I_T$ for the DW transfer. The magnetic bits are then read in parallel way for each 4-individual sections of the device by using the MOKE signal detection, basically mimicking a serial-in and parallel-out shift register. The bottom four panels show the MOKE signal at the centers of each section. The 'Bit 1' corresponds to the closest section to the recording current line. The results clearly demonstrate that all the arbitrary data bits are successively shifted to the next section for each clock. We confirm that the operation of the shift register is successfully accomplished more than a few tens of clocks.

In summary, we demonstrate the electric control of multiple DW motion by use of Pt/Co/Pt nanotracks with perpendicular magnetic anisotropy. By suppressing microstructural disorders in Pt/Co/Pt films, a low operation current for DW creep is attained and a four-bit operation is demonstrated. Our device prototype provides a step closer to an emerging magnetic domain-wall-based memory and logic device.

This work was supported by the National Research Foundation of Korea grant funded by the Korea government (2007-0056952, 2009-0084542). JCL and KHS were supported by the KIST Institutional Program, by the KRCF DRC program, and by the IT R&D program of MKE/KEIT (2009-F-004-01). KJK was supported by the Seoul R&BD program (10543). Two authors (K.-J.K. and J.-C.L) equally contributed to this work.

Figure Captions

**Figure 1**. (a) Polar MOKE microscope images of domain evolution patterns in Pt/Co/Pt film. Each domain image is obtained at the time denoted in the figure under magnetic field pulse (1.5 mT). (b) Secondary electron microscope (SEM) image of the device, composed of a nanotrack and three electrodes. (c) Schematic illustration of the experimental procedure.

**Figure 2**. (a) Line-scanned MOKE signals for current-driven DW motion. Data are obtained for each successive current pulses $I_T$ ($\pm 9.8 \times 10^6$ A/cm$^2$, 20 ms). Each plot is overlapped with the colored background, whose color contrast corresponds to the MOKE signal level—red for bit '1' (up) and blue for bit '0' (down). The arrows (orange and green) indicate the current polarity. The current-driven motions of the single DW (left) and the double DWs (right) are shown. (b) Line-scanned MOKE signals which are obtained for each clock, while recording arbitrary bits per clock. The bit sequences 1010 (left) and 1100 (right) are tested.

**Figure 3**. Operation timetable of four-bit shift register. The top panel indicates the recording current pulses $I_R$ with the bit sequence 00111011 repeatedly. The second panel shows the clock pulses $I_T$. The next four panels show the MOKE signal at the positions—2, 6, 10, and 14 μm away from the recording current line—corresponding to the centers the of 4-bit sections, respectively. The fluctuation of the MOKE signal at the 0th clock comes from the reset magnetic field pulse. For guide to the eyes, the first cycle is enlarged with shaded background.

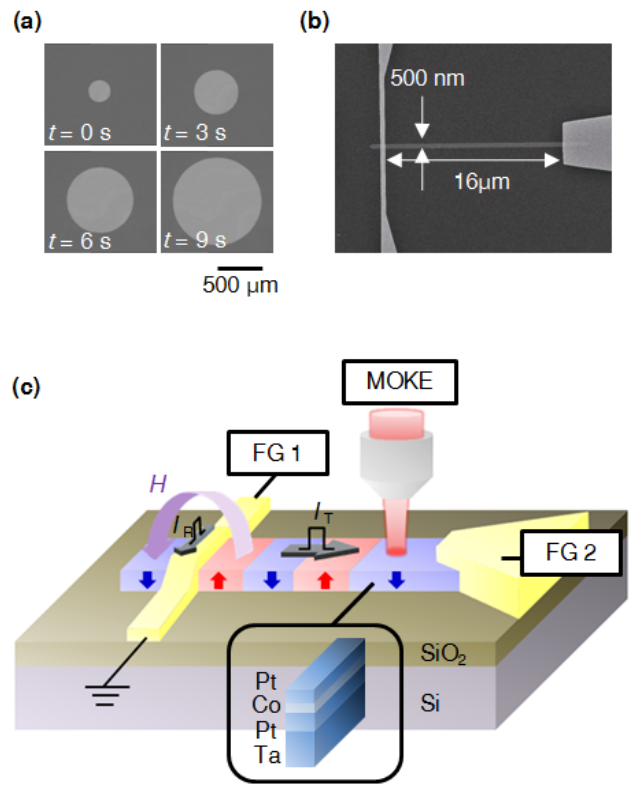

Figure 1

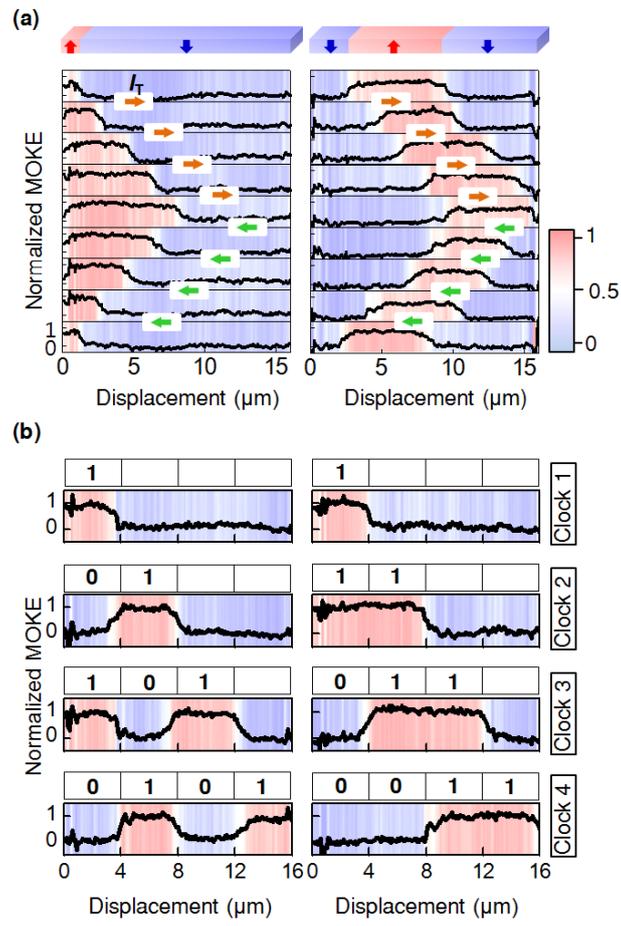

Figure 2

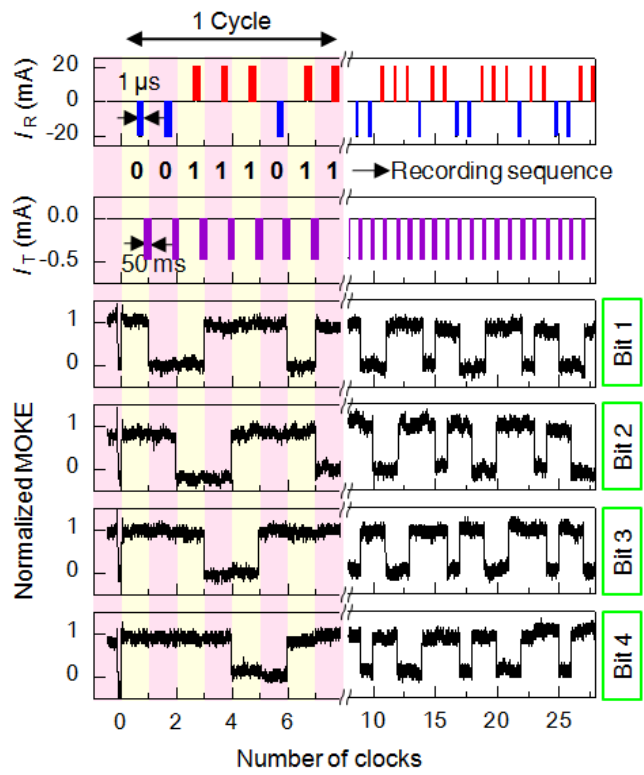

Figure 3